\documentclass[iicol]{sn-jnl}
\usepackage{graphicx}%
\usepackage{multirow}%
\usepackage{amsmath,amssymb,amsfonts}%
\usepackage{amsthm}%
\usepackage{mathrsfs}%
\usepackage[title]{appendix}%
\usepackage{xcolor}%
\usepackage{textcomp}%
\usepackage{manyfoot}%
\usepackage{booktabs}%
\usepackage{algorithm}%
\usepackage{algorithmicx}%
\usepackage{algpseudocode}%
\usepackage{listings}%
\usepackage[numbers]{natbib}
\usepackage{xcolor}
\usepackage{graphicx}
\usepackage{subcaption}

\theoremstyle{thmstyleone}%
\theoremstyle{thmstyletwo}%

\theoremstyle{thmstylethree}%

\begin{document}

\title[Article Title]{Interacting with Alexa: A Longitudinal Case Study of Voice Assistant Use in Down Syndrome}


\author*[1]{\fnm{Caterina} \sur{Senette}
%\orcidID{0000-0002-4411-7134}
}

\email{caterina.senette@iit.cnr.it}

\affil*[1]{\orgdiv{Istituto di Informatica e Telematica}, \orgname{Consiglio Nazionale delle Ricerche}, \orgaddress{\street{Via G. Moruzzi, 1}, \city{Pisa}, \postcode{56124}, \country{Italy} %\state{Italy} 
}}


\abstract{
As life expectancy for individuals with Down syndrome (DS) continues to increase, supporting cognitive engagement, everyday autonomy, and participation throughout adulthood has become an important research priority. Conversational voice assistants have emerged as promising assistive technologies for these purposes, yet evidence regarding their long-term use in naturalistic settings remains limited. This paper presents a longitudinal case study of the everyday use of Amazon Alexa by an adult with Down syndrome over nine months in a real domestic environment.
The study combines analyses of interaction logs, audio recordings, and qualitative observations to investigate usage patterns, interaction performance, communicative adaptation, and the integration of the voice assistant into everyday life.
Results show that Alexa became progressively embedded in daily routines, supporting entertainment, information access, and autonomy-related activities. Longitudinal analyses reveal stable interaction patterns, reduced retry behavior, and the emergence of effective communication strategies, while qualitative observations highlight increasing communicative engagement and the value of proactive reminders. The study also identifies important limitations of current commercial voice assistants, including reduced robustness to non-normative speech, limited conversational continuity, and poor contextual reasoning. Beyond providing rare ecologically valid evidence of long-term voice assistant use, the findings highlight design considerations that are relevant to the ongoing evolution of LLM-based conversational systems.
}




\maketitle
\section{Introduction and Motivation}\label{sec1}

Down syndrome (DS) is the most common genetic cause of intellectual disability, affecting approximately one in every 1,000 live births worldwide \cite{WHO2025DownSyndrome}. Advances in healthcare have substantially increased life expectancy over the past decades, with many individuals with DS in high-income countries now living well into adulthood and older age (exceeding 60 years in many countries) \cite{Bittles2004LifeExpectancy,Santoro2024Lancet}. Consequently, promoting cognitive health, autonomy, and participation throughout adulthood has become an increasingly important research priority. Although many individuals with DS acquire cognitive, social, and adaptive skills during childhood and adolescence through structured educational programs, the transition to adult life is often accompanied by reduced opportunities for cognitive stimulation, social participation, and lifelong learning \cite{Long2024AdulthoodDS}. This reduction in environmental enrichment, together with the high prevalence of age-related neurodegenerative changes associated with DS, may contribute to a gradual decline in adaptive functioning and everyday independence \cite{Santoro2024Lancet,deSola2021Surplus}. Consequently, there is a growing need for interventions capable of providing continuous cognitive stimulation while supporting communication, participation, and independent living throughout adulthood.

Accessible digital technologies have the potential to complement traditional support strategies by fostering cognitive engagement and promoting participation in everyday activities. However, conventional graphical user interfaces (GUIs) continue to present important accessibility barriers for individuals with DS. A growing body of research has shown that effective interaction with graphical interfaces often requires the simultaneous management of visual navigation, reading, memory, attention, and fine motor skills, creating significant obstacles to independent technology use \cite{Feng2008ComputerUsageYoung,Feng2010ComputerUsageChildren,Lazar2011AdultExpertUsers}. Similar challenges have been documented for web and touchscreen interfaces, where hierarchical navigation, icon interpretation, gesture-based interaction, and limited support for error recovery frequently increase cognitive load and reduce usability \cite{AlonsoVirgos2018WebRecommendations,AlonsoVirgos2018WCAG,AlonsoVirgos2020WebMethodology,Nacher2018TouchscreenChildren,Caliz2021TouchscreenAdults,MartinGutierrez2021TouchscreenGestures,Mendoza2015Learnability}. Consequently, several authors have argued for the development of alternative interaction paradigms capable of reducing reliance on visual navigation, symbolic interpretation, and precise motor actions \cite{Teran2016USATESTDOWN,Nascimento2021UsabilitySLR}.

\begin{figure*}[t]
    \centering
    \includegraphics[width=\textwidth]{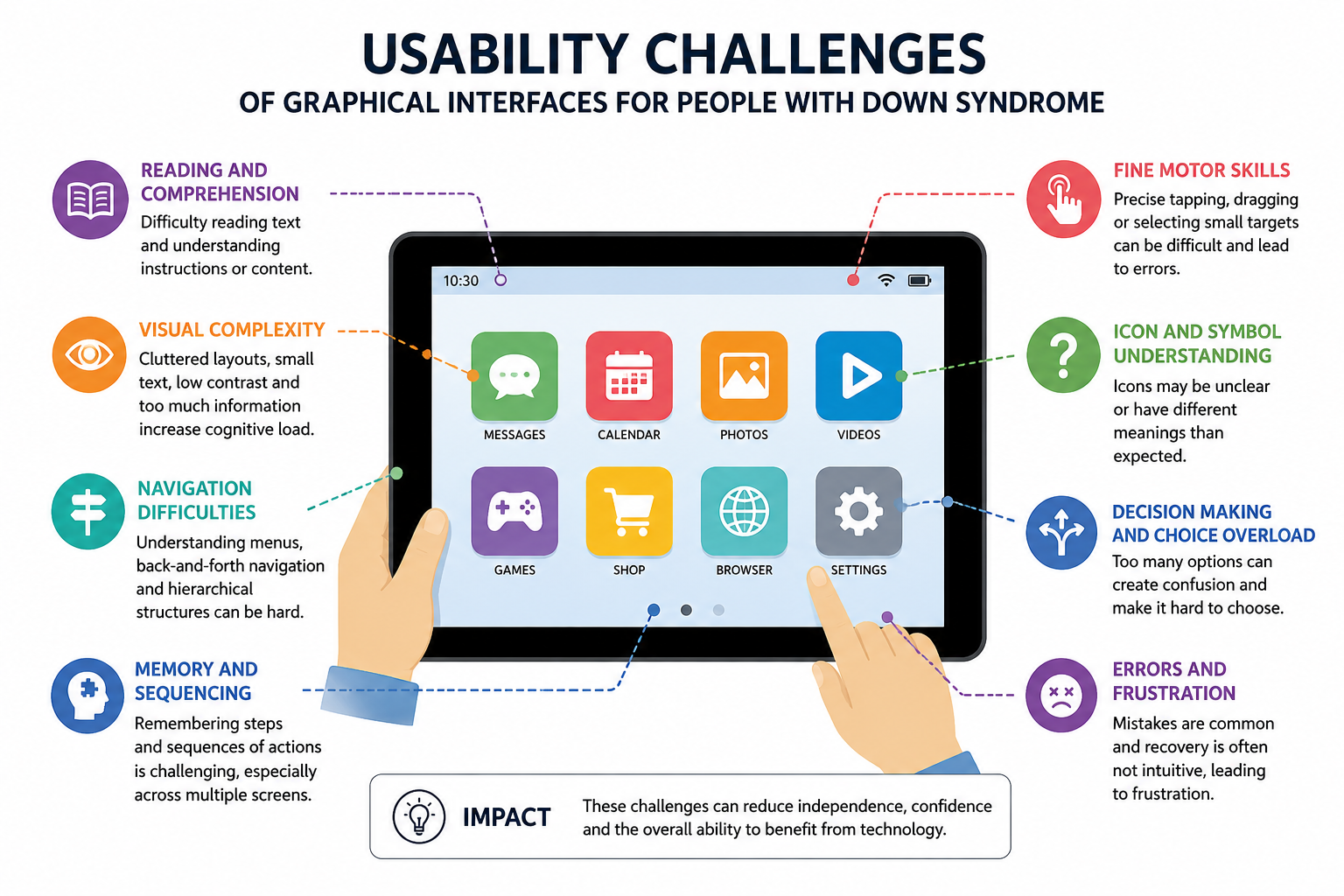}
    \caption{Conceptual overview of usability challenges commonly reported in the literature for graphical and touchscreen interfaces used by individuals with Down syndrome.}
    \label{fig:usability-challenges}
\end{figure*}

Fig.~\ref{fig:usability-challenges} summarizes the principal usability challenges consistently reported in the literature. Collectively, these barriers limit independent technology use and motivate the exploration of interaction paradigms that rely less on visual and motor demands.

Among the most promising alternatives are voice assistants (VAs) such as Amazon Alexa\footnote{\url{https://www.amazon.com/alexa}}, Google Assistant\footnote{\url{https://assistant.google.com/}}, and Apple Siri\footnote{\url{https://www.apple.com/siri/}}. By enabling access to digital services through natural language interaction, these systems reduce dependence on visual navigation, menu structures, and manual input while providing a more intuitive interaction modality. Their hands-free and conversational nature makes them particularly attractive for supporting individuals with cognitive and developmental disabilities in everyday activities.

Despite this potential, empirical evidence regarding the long-term use of commercial voice assistants by adults with Down syndrome remains scarce. Most available studies have focused on prototype systems, short-term deployments, or controlled experimental settings, leaving important questions unanswered regarding how commercially available conversational agents are appropriated, integrated into everyday routines, and adapted to over prolonged periods of use in natural domestic environments. Addressing these questions is essential for informing the design of future accessible, adaptive, and context-aware conversational systems capable of supporting communication, autonomy, and quality of life throughout adulthood.

\paragraph{Contribution}

%To address the limitations identified in prior work, this paper presents a longitudinal case study of a person with autism spectrum disorder (ASD) interacting with a commercial voice assistant (Amazon Alexa) over a period of six months in a real-world home environment.

%The main contributions of this work are as follows:

%\begin{itemize}
 %   \item We provide an in-depth, longitudinal account of how a voice assistant is appropriated and integrated into the daily routines of an individual with ASD, highlighting patterns of use, emerging practices, and changes over time.
    
 %   \item We identify key benefits of voice interaction in this context, including support for routine management, access to information, and engagement, as well as challenges related to speech recognition, interaction breakdowns, and system limitations.
    
 %   \item We offer qualitative insights into the user experience, shedding light on how voice assistants may influence autonomy, predictability, and interaction preferences in individuals with neurodevelopmental conditions.
    
  %  \item Based on our findings, we derive design implications for more inclusive and adaptive voice assistants, emphasizing the need for personalization, robustness to non-normative speech, and better alignment with users' cognitive and behavioral characteristics.
%\end{itemize}

This work contributes to Human--Computer Interaction (HCI) and Assistive Technology by presenting a nine-month longitudinal case study of the everyday use of a commercial conversational agent (Amazon Alexa) by an adult with Down syndrome in a natural domestic environment. Unlike previous short-term or laboratory-based investigations, it captures spontaneous interactions as they unfold in everyday life, providing ecologically valid evidence of technology appropriation, communicative adaptation, and routine formation.

The results show how a commercial voice assistant can become embedded in daily activities, supporting communication, entertainment, information access, routine management, and everyday autonomy, while simultaneously revealing persistent challenges related to non-normative speech, conversational continuity, and interaction breakdowns.

More broadly, the study extends current research on accessible conversational interfaces by deriving design implications for future voice assistants, emphasizing the need for greater robustness to diverse speech patterns, contextual awareness, adaptive dialogue management, and personalization according to users' cognitive and communicative characteristics.
\section{State of the Art}
Voice assistants (VAs) have attracted growing attention as accessible interaction technologies because their conversational and hands-free nature can reduce barriers associated with literacy, motor skills, and interface complexity. Accordingly, an increasing number of studies have investigated their potential to support independence, access to information, and participation in everyday activities for people with disabilities \cite{esquivel2024voice,tomasello2022impact}. Nevertheless, important challenges remain, including limited accessibility, insufficient personalization, and concerns regarding privacy and reliability \cite{volochtchuk2023voice}. Beyond interface usability, successful technology adoption has also been shown to depend on broader factors such as autonomy, self-confidence, caregiver support, and opportunities for meaningful participation in everyday life \cite{LussierDesrochers2017DigitalInclusion,Morris2023TechnologyUse,Johnson2024EasyUsingApps}.

Within the domain of cognitive and developmental disabilities, research on conversational agents has primarily focused on older adults and individuals with neurodegenerative conditions \cite{huq2022dialogue}. Comparatively less attention has been devoted to neurodevelopmental conditions, despite their distinctive communication profiles and interaction needs.

Among the studies conducted in neurodevelopmental contexts, Parvin et al.~\cite{parvin2022alexism} explored the use of Amazon Alexa to support oral hygiene routines in children with Autism Spectrum Disorder (ASD). In that work, the voice assistant primarily acted as a proactive instructional agent, providing reminders and guidance to encourage specific behaviors. While these findings demonstrated the potential of conversational agents as assistive technologies, interaction remained largely agent-driven and was evaluated in a structured intervention.

Research specifically targeting individuals with Down syndrome has followed a different trajectory. Most existing work has focused on educational applications, graphical interfaces, and augmentative and alternative communication (AAC) systems \cite{barbosa2018aac,boato2022virtual,buzzi2016learning,buzzi2019personalized}. These approaches emphasize structured, multimodal, and highly personalized interaction, but differ substantially from commercial voice assistants, which are designed as general-purpose speech-based interfaces with limited personalization.

Only recently have conversational voice user interfaces begun to receive attention as tools for supporting communication, engagement, and everyday activities for individuals with DS \cite{cibrian2025cvui}. Existing studies highlight the potential of conversational technologies to facilitate communication and daily assistance, while also identifying important limitations related to speech recognition, handling of non-normative speech, dialogue adaptability, and personalization. Furthermore, most investigations rely on prototype systems or short-term laboratory evaluations rather than commercially available assistants deployed in real domestic environments. More generally, research on the accessibility of commercial voice assistants has highlighted persistent issues related to interaction breakdowns, limited robustness to diverse speech patterns, and insufficient contextual reasoning \cite{masina2020accessibility}. These limitations are particularly relevant for individuals with DS, whose speech intelligibility, expressive language, and cognitive processing may differ substantially from those assumed during the design of mainstream conversational systems\cite{Christakis2019ParentChild,McMurray2019Lexical}.

A recent literature review by Cibrian et al.~\cite{Cibrian2025CVUIReview} provides the most comprehensive overview to date of conversational voice user interfaces for individuals with Down syndrome, analyzing 43 studies spanning speech recognition, conversational agents, therapeutic applications, and assistive technologies. The review highlights the promise of voice-based interaction for supporting communication, therapy, and everyday assistance, while emphasizing the importance of participatory design, adaptive dialogue strategies, and improvements in speech recognition technologies. At the same time, it identifies several limitations in the current literature, including the predominance of prototype systems, small-scale evaluations, and a strong focus on therapeutic applications and child populations. The authors further recommend extending conversational technologies to everyday activities, increasing support for adults with DS, and conducting more contextual investigations capable of informing the design of future conversational interfaces.

Taken together, the existing literature reveals two major gaps. First, most studies rely on short-term evaluations conducted in controlled settings, providing limited insight into how conversational agents are appropriated and integrated into everyday life over extended periods of use. Second, comparatively little attention has been devoted to adults with Down syndrome, despite their distinctive cognitive, linguistic, and communicative characteristics. %Third, commercially available voice assistants remain largely generic systems, offering limited personalization, adaptability, and robustness to non-normative speech. The present study addresses these limitations by investigating the long-term use of a commercially available voice assistant by an adult with Down syndrome in a natural domestic environment, combining interaction logs, audio recordings, and qualitative observations collected over approximately nine months.

Building on the literature recommendations and to overcome current limitations, the present study investigates the long-term use of a commercially available voice assistant (Amazon Alexa)\footnote{Alexa Echo Pop -- 1st Generation.} by an adult with DS in a real domestic environment. 
Moreover, differently from the literature, where the interaction remained largely agent-driven, the present study investigates the bidirectional interaction: the user not only responds to reminders but also actively initiates requests for information, entertainment, environmental control, and everyday support.
To guide this investigation, the following research questions were formulated:

\begin{itemize}
    \item \emph{RQ1: How is a conversational agent appropriated and incorporated into the everyday routines of a person with DS over an extended period of use?}

    \item \emph{RQ2: What interaction challenges emerge during long-term use of a voice assistant, and how are they addressed through user adaptation strategies?}

    \item \emph{RQ3: How does sustained interaction with a conversational agent contribute to communicative engagement, emotional regulation, routine formation, and everyday autonomy?}
\end{itemize}

\section{Methodology}

\begin{figure*}[h]
    \centering
    \includegraphics[width=\textwidth]{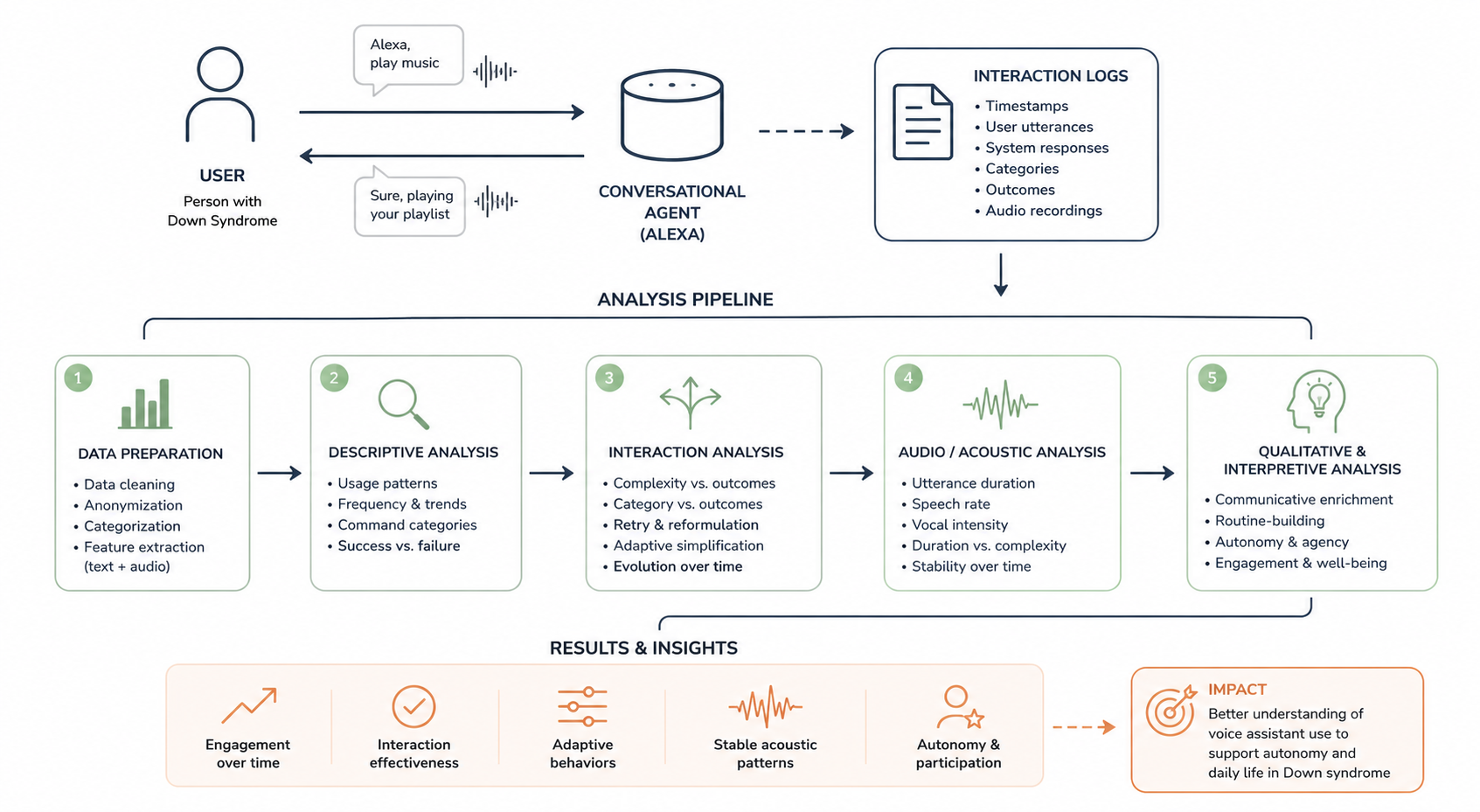}
    \caption{Overview of the system, data sources, analysis pipeline, and expected outcomes. The figure illustrates the interaction between the user and the conversational agent (Amazon Alexa), the collection of naturally occurring interaction logs and audio recordings, and the multi-level analysis conducted in the study, including descriptive, interaction, acoustic, and qualitative analyses.}
    \label{fig:alexa_pipeline}
\end{figure*}

\subsection{Research Design}

This work adopts a longitudinal case study, an approach particularly suited to investigating complex, context-dependent phenomena and generating in-depth insights into user experiences and technology appropriation. Consistent with established practices in HCI research, the objective is not statistical generalization but analytical generalization, namely the development of transferable insights grounded in rich empirical evidence.

By examining the everyday use of Amazon Alexa, this design makes it possible to capture not only initial interactions but also the evolution of routines, adaptation strategies, and sustained patterns of engagement that emerge through prolonged everyday use.

\subsection{Participant and Context}
The participant is an adult native Italian speaker diagnosed with DS; consequently, all interactions with Amazon Alexa were conducted in Italian. %The single-case design enabled an in-depth examination of how a commercial voice assistant was appropriated to support routines, leisure activities, information seeking, and autonomy over approximately nine months.

%The prolonged observation period provided a rare opportunity to investigate voice assistant use under natural conditions, capturing authentic patterns of engagement and adaptation that are difficult to observe in controlled experimental settings.

To protect participant confidentiality and minimize the risk of re-identification, potentially identifying information regarding the participant, the recruitment process, and the circumstances that enabled the study has been intentionally omitted. Accordingly, only contextual details necessary to interpret the findings are reported. When appropriate, observations from the participant are complemented by information provided by a caregiver concerning routines, behaviors, and perceived changes over time.

Although prolonged access to the study setting may introduce potential sources of bias, systematic data collection and consistent analytical procedures were employed throughout the study to enhance the rigor and credibility of the findings.

\subsection{Data Collection}

Data were collected over a nine-month period using two complementary data sources:

\begin{itemize}

\item \textbf{Interaction log analysis:} The study analyzed naturally occurring interaction logs collected from Amazon Alexa, focusing on patterns of engagement, frequency of interaction, recurring command categories, and interaction outcomes (e.g., successful or unsuccessful exchanges). The logs included both \textit{user-initiated} and \textit{agent-initiated} interactions serving different purposes. Most interactions consisted of user requests directed to the conversational agent, including information queries, entertainment requests, and routine tasks. In contrast, agent-initiated interactions involved proactive reminders, prompts, and notifications aimed at supporting everyday activities and autonomy, such as reminders to take medication, wear glasses before leaving home, go to the bathroom, or pay attention to household conditions and daily safety-related routines. This distinction provided insight into both spontaneous patterns of technology use and the assistive role of proactive voice-based interventions in everyday life.

\item \textbf{Contextual notes:} Contextual and observational notes were documented throughout the study to capture relevant details about everyday use, notable episodes, interaction breakdowns, and emergent behaviors observed within the domestic environment. Each observation was annotated with a timestamp, allowing it to be aligned with the corresponding interaction logs and audio recordings.

%\item \textbf{Interviews:} Semi-structured interviews were conducted with the participant and with the caregiver to gather reflections on the use of the voice assistant, perceived benefits, challenges, and changes in Interaction outcome over time.

\end{itemize}

Particular attention was given to identifying \emph{critical incidents}, including both successful interactions and breakdowns (e.g., failed speech recognition, misunderstandings), as these provide valuable insight into the interaction dynamics and user adaptation strategies.

\subsection{Analytical Approach}

The analysis combined quantitative, qualitative, and exploratory acoustic approaches. Interaction logs were examined to identify usage patterns, command categories, interaction outcomes, and longitudinal trends. Additional analyses explored linguistic complexity, retry behaviors following unsuccessful interactions, and selected acoustic characteristics of the recorded utterances. Finally, qualitative interpretation was used to contextualize the findings within everyday routines, autonomy-supporting practices, emotional regulation and communicative engagement.
\subsection{Ethical Considerations}

Given the involvement of a participant with a neurodevelopmental condition, this study adheres to ethical guidelines for research with vulnerable populations. Informed consent was obtained from the participant and their caregiver. All data were anonymized, and identifying details were removed or generalized.

Special care was taken to minimize intrusion into the participant’s daily life and to ensure that the study did not interfere with existing routines or support structures. Data collection methods were designed to be as unobtrusive as possible, prioritizing naturalistic observation over experimental manipulation.

\subsection{Limitations of the Methodological Approach}

As a single-case investigation involving one individual with DS, the findings are not intended to be statistically representative. Rather, the value of the work lies in the richness, duration, and ecological validity of the longitudinal data, which provide a unique opportunity to examine how a conversational agent is appropriated and integrated into everyday life over time. The resulting insights are intended to support analytical generalization and inform the design of inclusive conversational and assistive technologies. Future research should explore whether similar patterns emerge across individuals with different characteristics, support needs, and living contexts.
\section{Results and Discussion}\label{sec2}
\subsection{Usage Patterns}
This section addresses \emph{RQ1} by examining how the voice assistant was progressively incorporated into the participant's everyday routines.
\begin{figure*}[t!]
    \centering

    \begin{subfigure}[b]{0.48\textwidth}
        \centering
    \includegraphics[width=\linewidth]{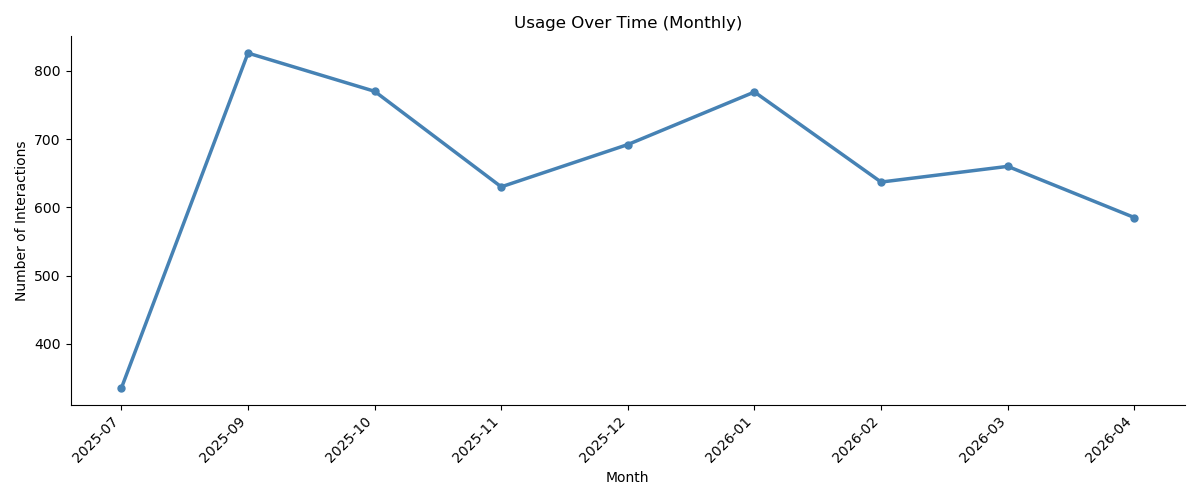}
        \caption{Interaction frequency over the observation period.}
        \label{fig:usage_time}
    \end{subfigure}
    \hfill
    \begin{subfigure}[b]{0.48\textwidth}
        \centering
        \includegraphics[width=\linewidth]{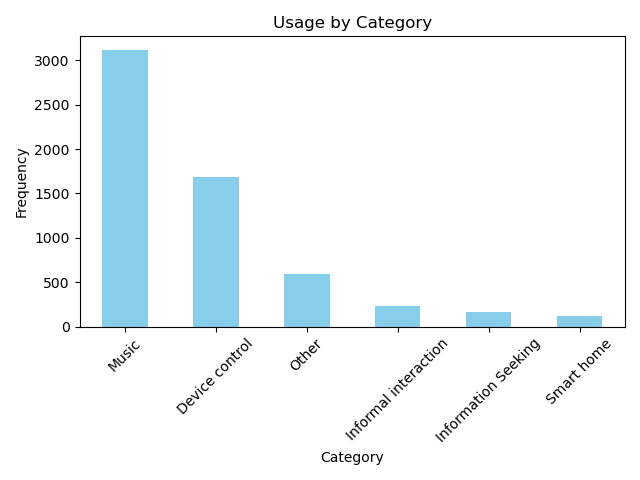}
        \caption{Distribution of interactions by category.}
        \label{fig:usage-cat}
    \end{subfigure}

    \vspace{0.5em}

    \begin{subfigure}[b]{0.48\textwidth}
        \centering
        \includegraphics[width=\linewidth]{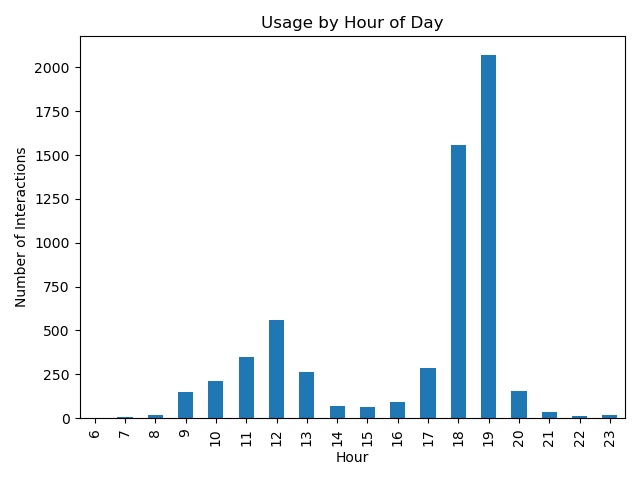}
        \caption{Distribution of interactions by hour of day.}
        \label{fig:usage_hour_day}
    \end{subfigure}
    \hfill
    \begin{subfigure}[b]{0.48\textwidth}
        \centering
        \includegraphics[width=\linewidth]{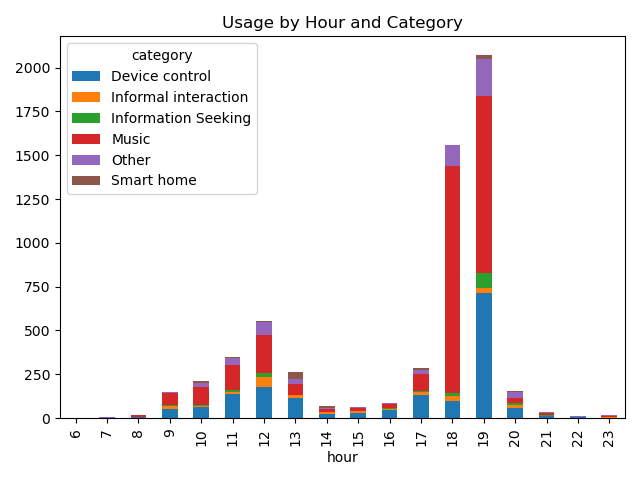}
        \caption{Interaction categories across hours of the day.}
        \label{fig:usage-hour-cat}
    \end{subfigure}

    \caption{Overview of usage patterns observed during the nine-month deployment. The figure reports (a) monthly interaction frequency, (b) distribution of interaction categories, (c) interaction frequency by hour of day, and (d) category-specific usage patterns across hours of the day.}
    \label{fig:overview_usage_patterns}
\end{figure*}
Fig.~\ref{fig:usage_time} provides an overview of Alexa usage throughout the observation period. As shown in the figure, interaction frequency remained relatively stable across the nine-month study, with monthly counts generally ranging between approximately 600 and 850 interactions. The absence of interactions in August 2025 reflects a period during which the device was intentionally switched off and therefore unavailable. Beyond this interruption, the data suggest sustained engagement with the voice assistant over time, with no evidence of progressive abandonment or declining use. Interestingly, after interruption, interaction frequency increased substantially when the device was reintroduced in September, reaching the highest monthly usage observed during the study. Although no causal conclusions can be drawn, this pattern may reflect renewed interest and engagement following the period of non-use.

The distribution of interaction categories is reported in Fig.~\ref{fig:usage-cat}. Music-related requests constituted the largest category, followed by \textit{device-control} interactions and a broad set of miscellaneous activities grouped under the \textit{other} category, such as residual interactions that did not fit the predefined classification scheme, including ambiguous or incomplete utterances, fragmentary requests, residual control commands, and miscellaneous conversational expressions. 
\textit{Information-seeking}, \textit{smart-home} commands, and \textit {informal interactions} represented a comparatively small proportion of the overall activity. This distribution indicates that the voice assistant was primarily used for entertainment and practical everyday support.

Temporal usage patterns are illustrated in Fig.~\ref{fig:usage_hour_day}. Interactions were concentrated around specific periods of the day, with a pronounced peak during the evening hours (18:00–19:00) and a secondary peak around midday. Activity was considerably lower during the early morning and late evening, suggesting that voice assistant use was closely associated with recurring daily routines.

Finally, Fig.~\ref{fig:usage-hour-cat} combines temporal and categorical information. The evening peak was largely driven by music requests and device-control commands, while the remaining categories contributed only marginally to overall usage. 
The consistency of these patterns suggests that Alexa evolved from a simple technological tool into a stable component of the participant's everyday life, supporting both recreational and functional activities. This provides an initial answer to \emph{RQ1} regarding the long-term appropriation of conversational agents.
\subsection{Interaction Success, Failures, and Adaptation Strategies}

To address \emph{RQ2}, we investigated the challenges associated with long-term voice interaction and the user adaptation strategies that emerged over time. Specifically, we analyzed interaction outcomes across usage categories, linguistic complexity levels, and longitudinal trends.
Linguistic complexity refers to the structural formulation of the participant's utterances and was operationalized into three categories: two-word phrases, short phrases, and full sentences.
Interaction outcome refers to the result of a user utterance and is classified as complete command (the user's intent was successfully fulfilled), simple command (the interaction was successful but required a simplified command), or failure (the user's intent was not achieved).

\begin{figure*}[t]
    \centering
    \begin{subfigure}[b]{0.48\textwidth}
        \centering
        \includegraphics[width=\linewidth]{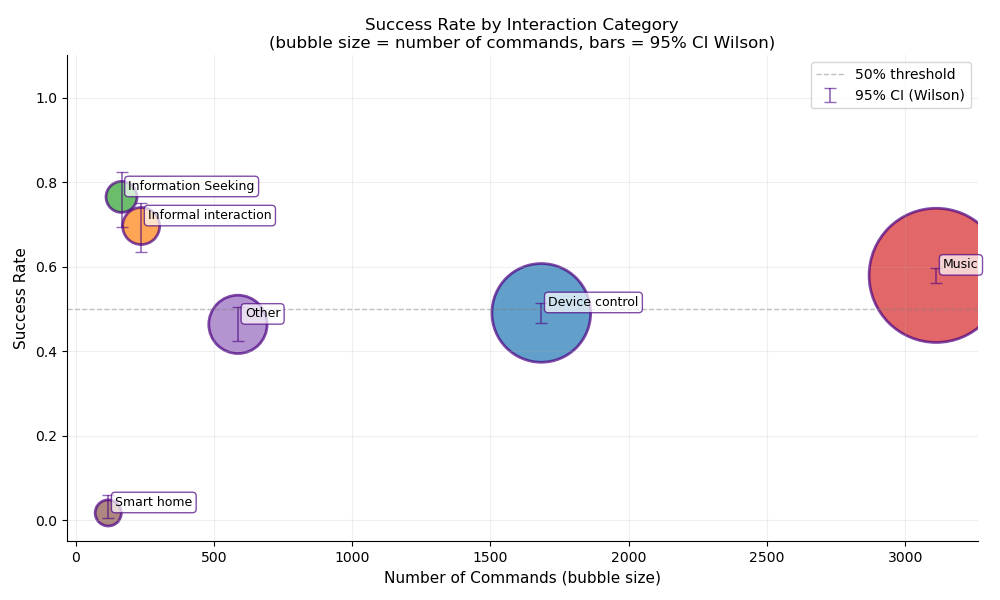}
        \caption{Success rate by interaction category.}
        \label{fig:success_category}
    \end{subfigure}
    \hfill
    \begin{subfigure}[b]{0.48\textwidth}
        \centering \includegraphics[width=\linewidth]{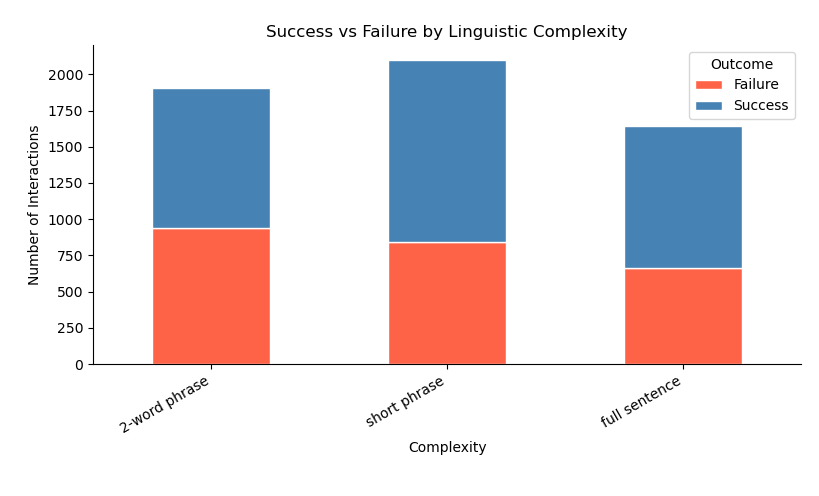}
        \caption{Success rate by linguistic complexity.}
        \label{fig:success_complexity}
    \end{subfigure}

    \caption{Interaction outcomes across categories and linguistic complexity levels.}
    \label{fig:success_analysis}
\end{figure*}

Fig.~\ref{fig:success_category} shows that success rates varied across interaction categories. \textit{Information-seeking} and \textit{informal interactions} achieved high success rates (influenced by the relatively small number of interactions belonging to these categories), whereas \textit{music} requests and \textit{device-control} commands, despite accounting for the majority of interactions, showed moderate performance. Wilson 95\% confidence intervals confirmed that the observed differences in success rates across interaction categories were generally robust, while also highlighting the greater uncertainty associated with less frequently used categories such as \textit{other} and \textit{smart home}. %Informal interactions and information-seeking requests achieved the highest success rates, whereas smart-home commands were associated with the lowest performance, suggesting that interaction breakdowns were influenced by the specific demands of different tasks.  Information-seeking and informal interactions achieved high success rates, whereas music requests and device-control commands—despite accounting for the majority of interactions—showed only moderate performance.

 Regarding successful and unsuccessful interactions across linguistic complexity levels, Fig.~\ref{fig:success_complexity} shows that short phrases constitute the most common interaction pattern and exhibit the highest number of successful exchanges. While success rates are broadly comparable across the three categories, the results indicate that increasing linguistic complexity did not substantially compromise interaction performance, suggesting the emergence of effective communication strategies over time.

\begin{figure*}[t]
    \centering

    \begin{subfigure}[t]{0.45\textwidth}
        \centering
    \includegraphics[width=\linewidth]{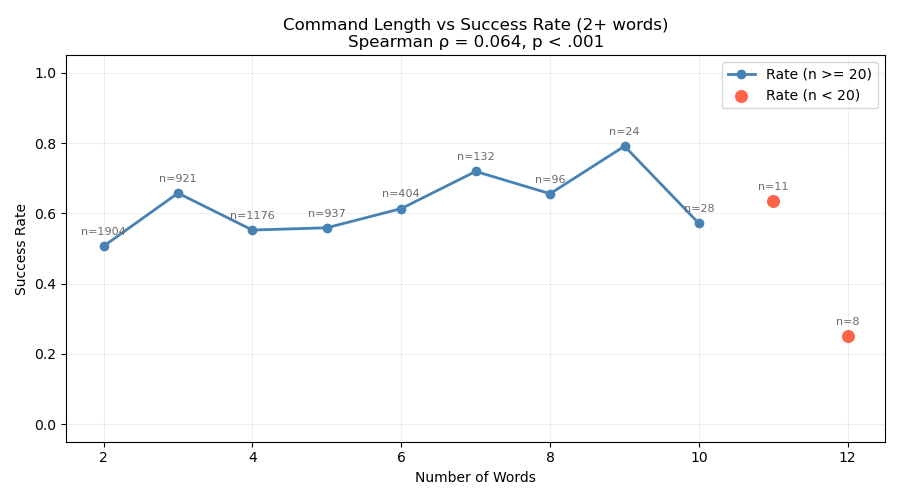}
        \caption{Success rate by command length.}
        \label{fig:successBy-command-lenght}
    \end{subfigure}
    \hfill
    \begin{subfigure}[t]{0.5\textwidth}
        \centering
        \includegraphics[width=\linewidth]{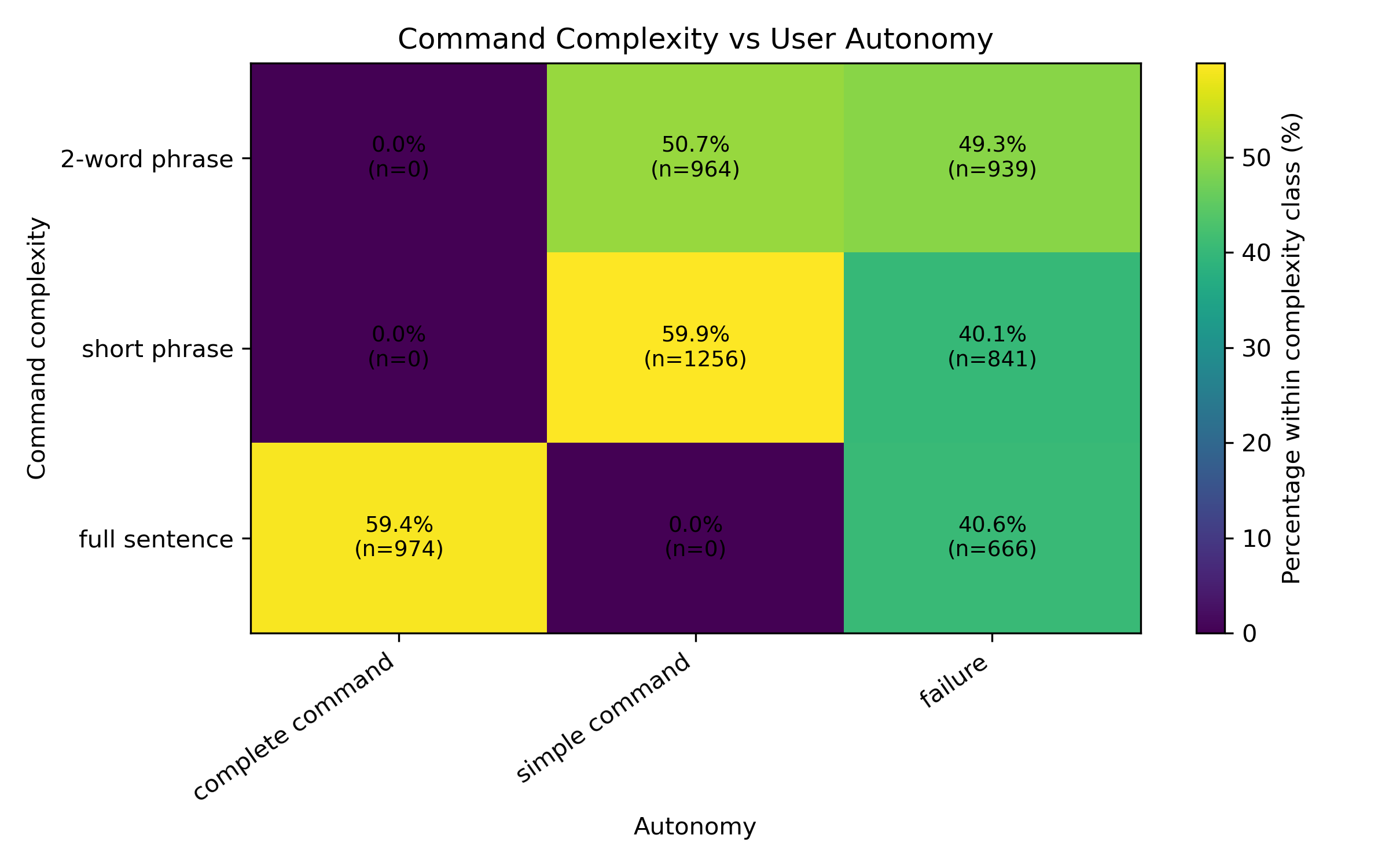}
        \caption{linguistic complexity vs interaction outcome.}
        \label{fig:heatmap}
    \end{subfigure}

    \caption{Relationship between command characteristics and interaction outcomes.}
    \label{fig:combined_results}
\end{figure*}

Fig.~\ref{fig:successBy-command-lenght} further explores the relationship between utterance length and interaction outcomes. Success rates remained relatively stable across commands ranging from two to eleven words, with moderate variability but no clear monotonic trend. The apparent variability observed for longer utterances is associated with a limited number of observations, as indicated by the reported sample sizes (\emph{n}); categories represented by fewer than 20 utterances are highlighted in red. A Spearman rank correlation revealed a statistically significant but negligible association between command length and interaction success ($\rho = 0.064$, $p < .001$), indicating that utterance length alone was not a meaningful predictor of successful interaction. % except in the case of the longest requests.
%%vedere se sostituire con questa heatmap

To illustrate the relationship between linguistic complexity and interaction success, Fig.~\ref{fig:heatmap} highlights that full sentences were predominantly associated with completely successful commands (59.4\%), indicating that more articulated utterances often enabled fully autonomous interactions. In contrast, short phrases were mainly classified as simple successful commands (59.9\%), whereas two-word phrases exhibited an almost equal distribution between successful and failed interactions (50.7\% vs.\ 49.3\%). These findings suggest that interaction success was not solely determined by linguistic complexity. Rather, very short utterances appeared to be more susceptible to ambiguity and recognition errors, whereas more structured commands enabled richer and more reliable interactions when successfully interpreted by the conversational agent.

\begin{figure}[t]
    \centering
    \includegraphics[width=1.0\linewidth]{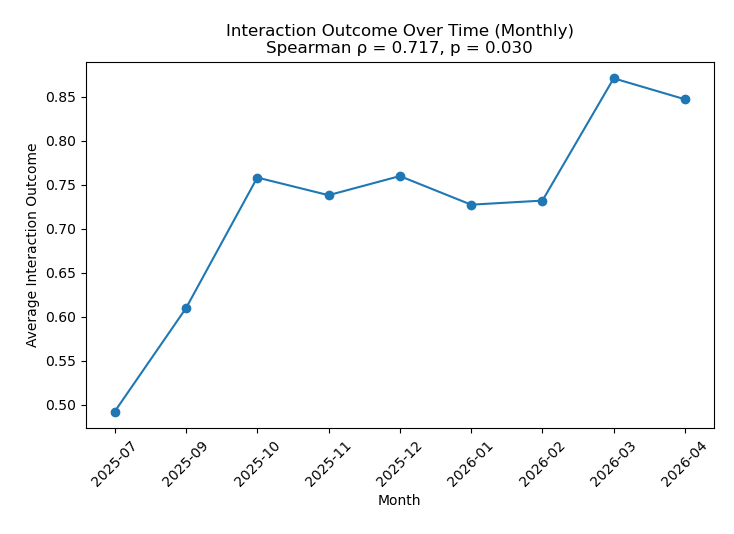}
    \caption{Monthly average interaction outcome score over the observation period. }
    \label{fig:autonomy_monthly}
\end{figure}

Longitudinal trends provide further evidence of user adaptation. As shown in Fig.~\ref{fig:autonomy_monthly}, interaction outcomes improved rapidly during the first months of use, followed by a prolonged period of stable performance and a further increase during the final phase of the study. A Spearman rank correlation revealed a significant positive association between time and interaction outcome ($\rho = 0.717$, $p = .030$), indicating a gradual improvement in interaction effectiveness over the observation period. These findings suggest that continued exposure to the conversational agent supported the development of increasingly effective interaction strategies.

Taken together, these findings indicate that successful interaction was shaped by both task characteristics and linguistic formulation. At the same time, the longitudinal stability of interaction outcome measures suggests that prolonged exposure to the voice assistant supported the emergence of effective communication strategies and a progressively consolidated interaction style.

\subsection{Exploratory Acoustic Findings}
The exploratory acoustic analyses further contribute to \emph{RQ2} by providing complementary evidence of communicative adaptation over time.
% Acoustic findings figure
\begin{figure*}[t]
\centering

\begin{subfigure}[t]{0.48\textwidth}
\centering
\includegraphics[width=0.9\linewidth]{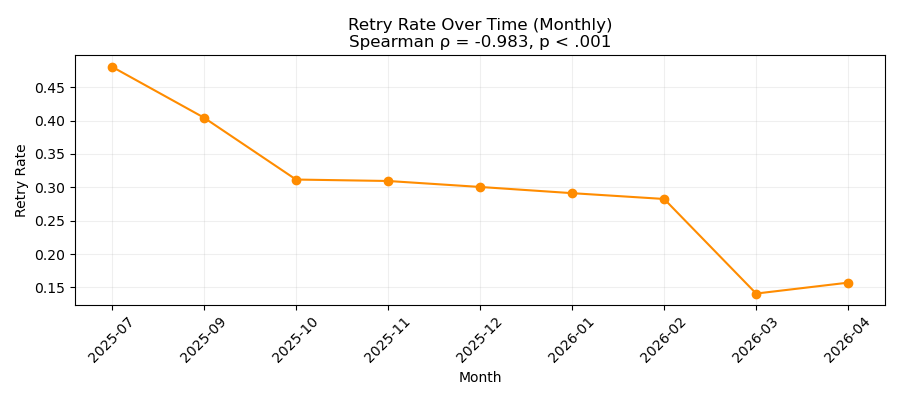}
\caption{Retry rate over time}
\label{fig:retry-rate}
\end{subfigure}
\hfill
\begin{subfigure}[t]{0.48\textwidth}
\centering
\includegraphics[width=0.9\linewidth]{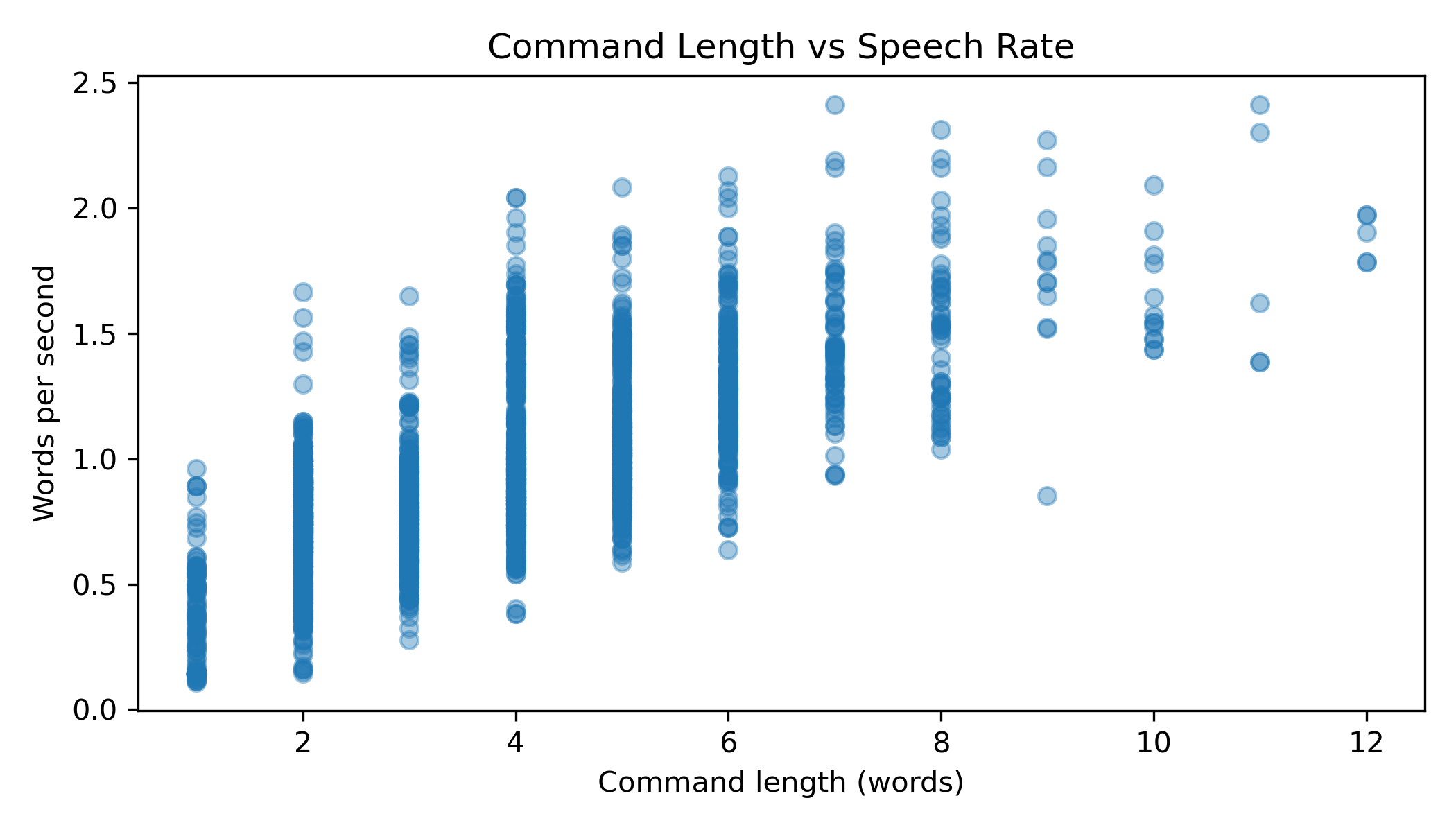}
\caption{Command length vs speech rate}
\label{fig:command-lenght}
\end{subfigure}

%\vspace{0.5em}

\begin{subfigure}[t]{0.48\textwidth}
\centering
\includegraphics[width=0.9\linewidth]{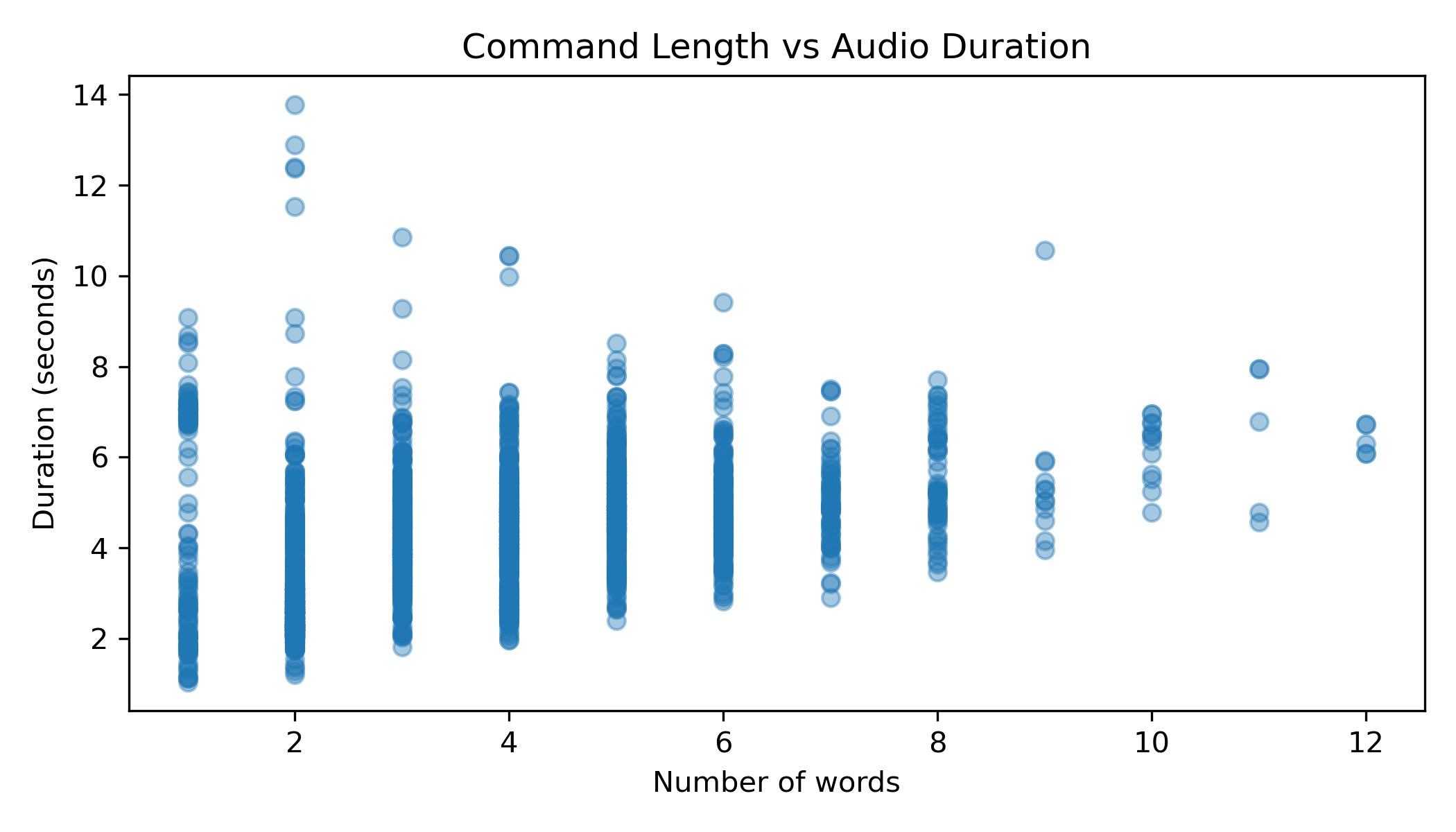}
\caption{Command length vs audio duration}
\label{fig:lenght-duration}
\end{subfigure}
\hfill
\begin{subfigure}[t]{0.48\textwidth}
\centering
\includegraphics[width=0.9\linewidth]{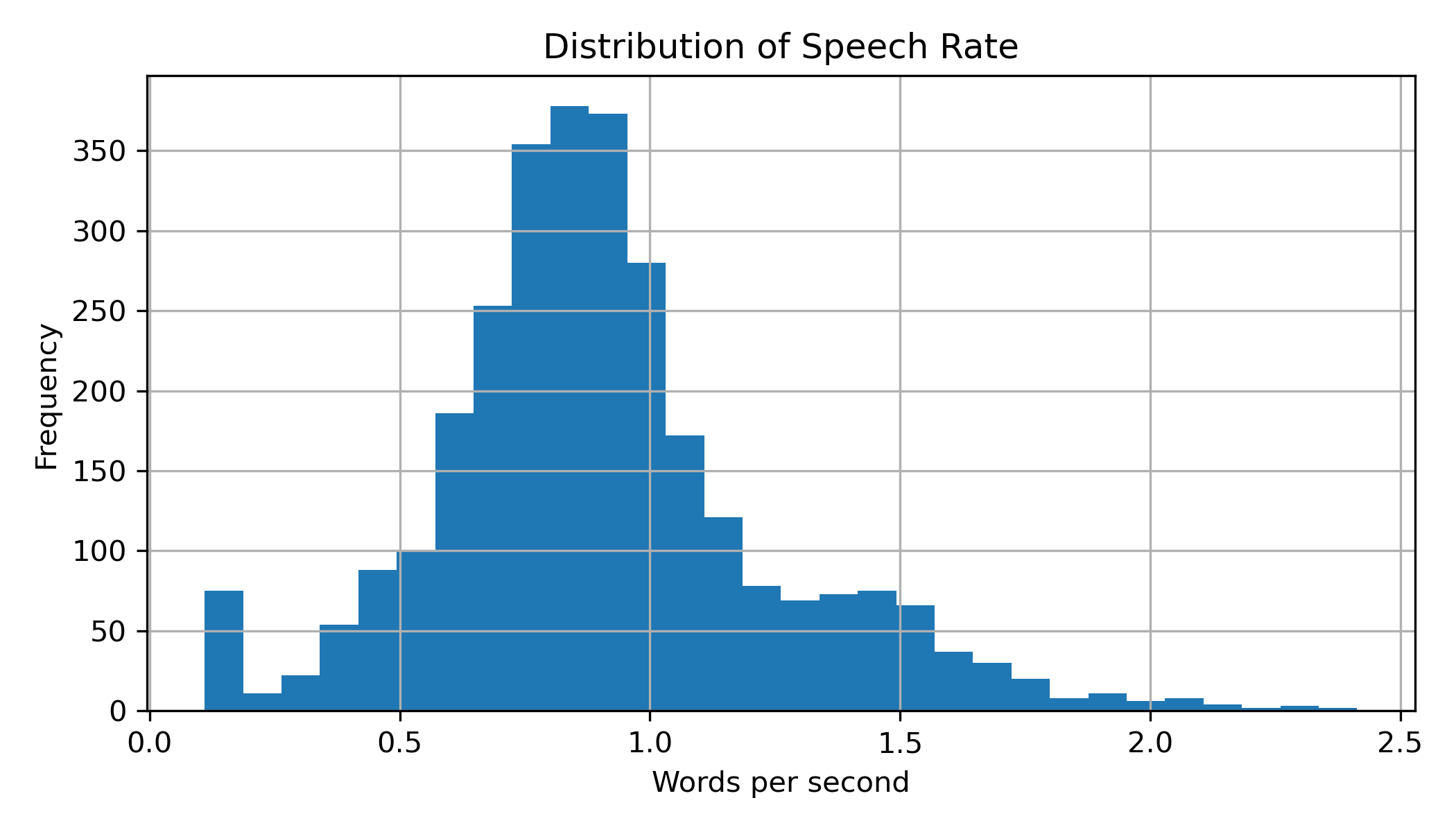}
\caption{Speech rate distribution}
\label{fig:rate-distribution}
\end{subfigure}

\caption{Exploratory acoustic analyses of long-term interaction with Alexa. Panel (a) reports the evolution of retry behavior over time, suggesting a reduction in command reformulations following unsuccessful interactions. Panels (b) and (c) show the relationships between command length, speech rate, and utterance duration, providing insight into communication strategies adopted during interaction. Panel (d) reports the distribution of speech rates across all recorded utterances, highlighting the relative stability of the participant's communication rhythm.}
\label{fig}
\end{figure*}

We conducted an exploratory examination of the audio recordings associated with the participant's voice commands\footnote{The audio recordings collected during this study, after complete anonymization and removal of all personally identifiable information, will be released as an open research resource to complement existing speech datasets for individuals with Down syndrome, supporting future research on more robust and inclusive automatic speech recognition systems for non-normative speech.}. The goal was not to assess speech quality or clinical language outcomes, but rather to investigate whether prolonged interaction with the voice assistant was associated with the emergence of stable communication patterns and interaction routines.

The clearest evidence of interaction stabilization is provided by Fig.~\ref{fig:retry-rate}. The proportion of repeated commands following unsuccessful interactions generally decreased throughout the observation period, dropping from approximately 20\% during the early months to below 10\% in the final phase of the study. This is also confirmed by statistics that highlight a very strong negative correlation between retry rate and time ($\rho = -0.983$, $p < .001$), indicating a substantial reduction in command reformulations throughout the observation period and providing strong evidence of progressive user adaptation to the conversational agent that reduced the need for command reformulation.

Additional insights emerge from the \emph{Command length vs speech rate} analysis (Fig.~\ref{fig:command-lenght}). Longer utterances were generally produced at slightly higher speech rates, suggesting that increased command length reflected richer linguistic formulation rather than slower speech production. This pattern is consistent with increasingly efficient interaction strategies.

Considering utterance duration, Fig.~\ref{fig:lenght-duration} confirms the expected positive relationship between command length and speaking time. While longer commands naturally required more time to utter, considerable variability remained among commands of comparable length, suggesting that the participant adopted different pacing and speech production strategies across interactions.

Finally, Fig.~\ref{fig:rate-distribution} shows that most utterances were produced within a relatively narrow speech-rate range, indicating a fairly consistent communication rhythm across the observation period.

Overall, these acoustic patterns are consistent with the emergence of stable and functional voice-based interaction routines. Importantly, they should not be interpreted as indicators of speech improvement. Rather, they provide exploratory evidence of communicative adaptation and increasing interaction performance during prolonged engagement with the conversational agent.

\subsection{Qualitative Insights}
The following qualitative findings primarily address \emph{RQ3} by contextualizing how prolonged interaction influenced communicative engagement and everyday autonomy.
\subsubsection{From Content Consumption to On-Demand Access}

Field observations suggest that the participant's understanding of Alexa evolved substantially over time. %Initially, interactions were relatively infrequent and exploratory. 
Before adopting the voice assistant, the participant had attempted to use a tablet during her leisure time to search for information of personal interest and access her favorite music. However, the interaction proved frustrating because of the usability challenges associated with graphical interfaces discussed in Section~\ref{sec1}. Difficulties related to visual navigation, menu structures, icon interpretation, and the cognitive demands of touchscreen interaction ultimately led her to abandon the device. As a result, her primary leisure activity became listening to music through an MP3 player. While this solution was familiar and easy to use, it also reinforced the perception of digital content as finite, static, and dependent on external updates, an expectation that was reflected in the participant's initial interactions with Alexa. Over time, however, the participant progressively understood that Alexa provided continuous access to a virtually unlimited catalog of music, information, and services. This shift was accompanied by increased engagement and more frequent use, suggesting a gradual appropriation of the technology as an always-available resource rather than a fixed-content device.

\subsubsection{Communicative Adaptation and Vocabulary Constraints}

Several observed behaviors help explain the interaction patterns identified in the quantitative analyses. During the initial stages of use, the participant experienced difficulties with the \textit{wake word} ``Alexa'' and repeatedly employed the simplified form ``Alessa''. Over time, the participant appeared to realize that perfect pronunciation was not required for successful interaction, leading to more fluent and confident use of the system.

At the same time, some commands remained persistently challenging. In particular, terms such as ``notification'' and ``stop'' were not part of the participant's everyday vocabulary. While the function associated with stopping media playback could often be achieved through alternative expressions (e.g., equivalents of ``basta'' or ``salta (skip)''), the notification functionality depended on a specific lexical item. This observation helps explain the high failure rate associated with notification-related commands and suggests that interaction difficulties were sometimes driven by vocabulary requirements imposed by the interface rather than by the underlying task itself.

Another recurring source of interaction breakdowns was related to environmental conditions rather than speech production. The participant showed a strong preference for high playback volumes and frequently increased Alexa's output level. Consequently, commands such as ``Alexa, stop'' were occasionally not recognized because they were issued while loud audio was being played, creating adverse acoustic conditions for speech recognition.

\subsubsection{Personification, Engagement, and Social Interaction}

Qualitative evidence further revealed a progressive tendency to personify the voice assistant. The participant frequently referred to the device as ``my Alexa'' and expressed concern when other family members interacted with it, believing that their actions could alter Alexa's behavior or preferences. This perception suggests that the device was not viewed merely as a technological tool but as a personalized entity associated with the participant's own routines and experiences.

Over time, interactions also expanded beyond purely instrumental requests. After an initial familiarization period, the participant began greeting Alexa in the morning, saying goodnight before going to sleep, asking how she was doing, and posing questions motivated by curiosity rather than immediate practical needs. Interestingly, these more socially oriented interactions occurred primarily when the participant was alone with the device. Such behaviors indicate that the conversational agent increasingly served not only as a functional assistant but also as a source of engagement and companionship, highlighting dimensions of use that are difficult to capture through interaction logs alone.
%\subsubsection{Communicative Engagement and Behavioral Indicators of Emotional Regulation}

%Beyond functional interaction, conversational agents may also influence users' communicative engagement, perceived companionship, and the regulation of everyday activities. Previous studies have suggested that voice assistants can contribute to wellbeing by supporting routines, reducing feelings of isolation, and providing predictable forms of interaction. However, little is known about how these aspects emerge during prolonged everyday use by individuals with Down syndrome.

Rather than attempting to infer emotional states directly, we analyzed interaction logs together with longitudinal field observations to identify observable behavioral indicators of communicative engagement and emotional regulation. Interactions were classified through a rule-based coding procedure based on lexical patterns and communicative intent, followed by manual review of all relational and ambiguous utterances to ensure coding consistency.
\begin{figure}[t]
    \centering
    \includegraphics[width=1.0\linewidth]{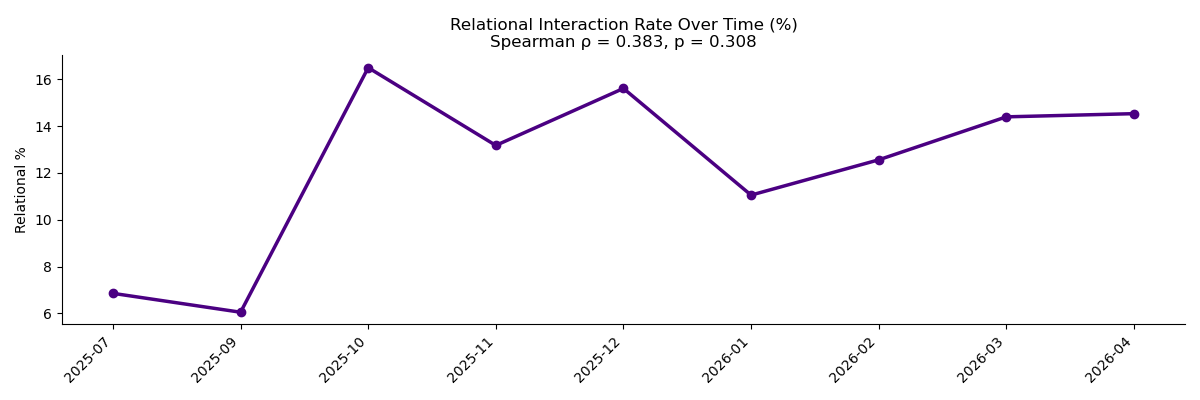}
    \caption{Monthly proportion of relational interactions. Although the proportion fluctuated throughout the observation period, relational exchanges consistently represented a stable component of the participant's interactions with the conversational agent.}
    \label{fig:relational_rate}
\end{figure}
\begin{table}[t]
\centering
\caption{Behavioral indicators of communicative engagement identified during the longitudinal study.}
\label{tab:relational_behaviour}
\begin{tabular}{p{1.8cm}p{1.2cm}p{3.5cm}}
\hline
\textbf{Behavior} & \textbf{Frequency} & \textbf{Interpretation} \\
\hline
\hline
Social conversation & 565 &
Use of Alexa beyond functional requests to satisfy curiosity and engage in spontaneous dialogue. \\

Farewell & 69 &
Conversational closure reflecting the incorporation of Alexa into everyday routines. \\

Greeting & 48 &
Spontaneous initiation of social interaction with the conversational agent. \\

Curiosity & 32 &
Questions extending beyond immediate task completion, suggesting exploratory use of the system. \\

Politeness & 12 &
Courtesy expressions directed toward the conversational agent. \\

Hostility & 0 &
No hostile or aggressive interactions were observed throughout the study. \\
\hline
\end{tabular}
\end{table}
Table~\ref{tab:relational_behaviour} summarizes the main behavioral indicators of communicative engagement observed during the study. Although task-oriented interactions accounted for the majority of the participant's use of Alexa, social conversations represented the most frequent behavior within the subset of relational interactions, followed by greetings, farewells, curiosity-driven exchanges, and politeness expressions. Together, these behaviors indicate that the participant's interaction with the conversational agent progressively extended beyond purely functional requests and became integrated into everyday social routines.

Fig.~\ref{fig:relational_rate} shows that relational interactions consistently accounted for approximately 10--16\% of monthly interactions throughout the observation period. Although a slight positive trend was observed, the association between time and the proportion of relational interactions was not statistically significant (Spearman $\rho = 0.383$, $p = .308$), suggesting that socially oriented exchanges remained a stable component of the participant's communication style rather than progressively replacing task-oriented interactions.

Importantly, these behaviors should not be interpreted as direct evidence of emotional regulation. Instead, they represent observable behavioral indicators of communicative engagement and of the participant's increasing tendency to incorporate the conversational agent into everyday life. 
%The qualitative observations presented in the following sections further contextualize these findings, showing how repeated social interactions, personification, and proactive reminders contributed to the development of a meaningful long-term relationship with the voice assistant.
\subsubsection{Agent-Initiated Interactions and Routine Support}
%%tabella
\begin{table*}[t]
\centering
\caption{Observed changes associated with the long-term adoption of the conversational agent. The table summarizes the main transformations documented through interaction logs, qualitative observations, and contextual evidence.}
\label{tab:before_after}
\renewcommand{\arraystretch}{1.25}
\begin{tabular}{p{2.5cm}p{4.2cm}p{4.8cm}p{2.8cm}}
\hline
\textbf{Aspect} &
\textbf{Before Alexa} &
\textbf{After Alexa} &
\textbf{Evidence} \\
\hline
\hline
Access to digital content &
Tablet interaction was abandoned because of the usability challenges associated with graphical interfaces. Music was primarily accessed through an MP3 player with fixed content. &
Natural-language interaction enabled immediate access to music, information, and entertainment without relying on graphical interfaces. Digital content was progressively perceived as dynamic and virtually unlimited. &
Field observations, interaction logs \\

Leisure activities &
Listening to locally stored music represented the participant's main leisure activity. &
Music streaming became the predominant use of Alexa, complemented by spontaneous information seeking and curiosity-driven interactions. &
Interaction logs, qualitative observations \\

Medication adherence &
Medication reminders were primarily provided by the caregiver, with an estimated adherence of approximately 50\%. &
Agent-initiated reminders resulted in approximately 95\% adherence overall (1,280/1,350), reaching 100\% adherence for medication reminders (270/270). &
Qualitative observations \\

Communication style &
Technology was perceived primarily as a functional tool. &
Interaction progressively evolved beyond task execution, including greetings, farewells, social conversations, politeness expressions, and anthropomorphic behaviors (``Alexa mia''). &
Interaction logs, qualitative observations \\

Interaction efficiency &
No prior experience with conversational interfaces. &
Progressive reduction in retry behavior, improved interaction outcomes, and stable communication strategies emerged throughout the observation period. &
Interaction logs, acoustic analyses \\

Routine support &
Daily routines relied predominantly on caregiver support. &
Alexa became integrated into everyday routines through proactive reminders, scheduled prompts, and repeated interactions supporting autonomy-related activities. &
Interaction logs, qualitative observations \\

\hline
\end{tabular}
\end{table*}

Another aspect of the participant's interaction with Alexa concerned agent-initiated prompts. Beyond responding to user requests, the voice assistant proactively delivered reminders at specific moments of the day, consistently addressing the participant by name. These prompts supported everyday activities, including self-care, household routines, and other autonomy-related tasks.

Observational data indicate that this proactive support was positively received and became an appreciated component of the participant's daily routine. The participant frequently responded verbally to Alexa's prompts before carrying out the suggested activity. Overall, approximately 95\% (1,280 out of 1,350) of agent-initiated prompts (in nine months) requiring user action were followed by successful task completion, typically with a high degree of motivation and engagement. Notably, adherence was complete for medication reminders (270 episodes), with the participant taking the medication in every observed case.
This pattern suggests that the reminders functioned not only as memory aids but also as sources of encouragement and behavioral activation.

An additional factor contributing to the effectiveness of these interactions was the combination of predictable and variable reminders. Some prompts were delivered regularly at fixed times and became integrated into established daily routines. Others were introduced less predictably and varied across days, reducing habituation effects and helping maintain the participant's attention and interest. The observed responses suggest that this balance between routine and novelty contributed to sustaining engagement with the system over time.

Table~\ref{tab:before_after} synthesizes the principal changes observed throughout the study by contrasting the participant's technology use before and after the introduction of Alexa. Rather than simply documenting interaction patterns, the table highlights how the conversational agent transformed several aspects of everyday life, including access to digital content, leisure activities, routine management, communicative engagement, and medication adherence. 
%These observations highlight the potential of conversational agents to support everyday autonomy not only through user-initiated requests but also through appropriately designed proactive interventions embedded within daily routines.

\subsection{Design Implications for Accessible Voice Assistants}
The findings addressing \emph{RQ1--RQ3} extend beyond the specific case investigated, providing empirical evidence that informs the design of future accessible conversational interfaces. The observed interaction patterns, adaptation strategies, and communication challenges highlight several opportunities for improving the robustness, contextual awareness, personalization, and autonomy-support capabilities of next-generation voice assistants.

\begin{figure*}[h]
    \centering
    \includegraphics[width=\textwidth]{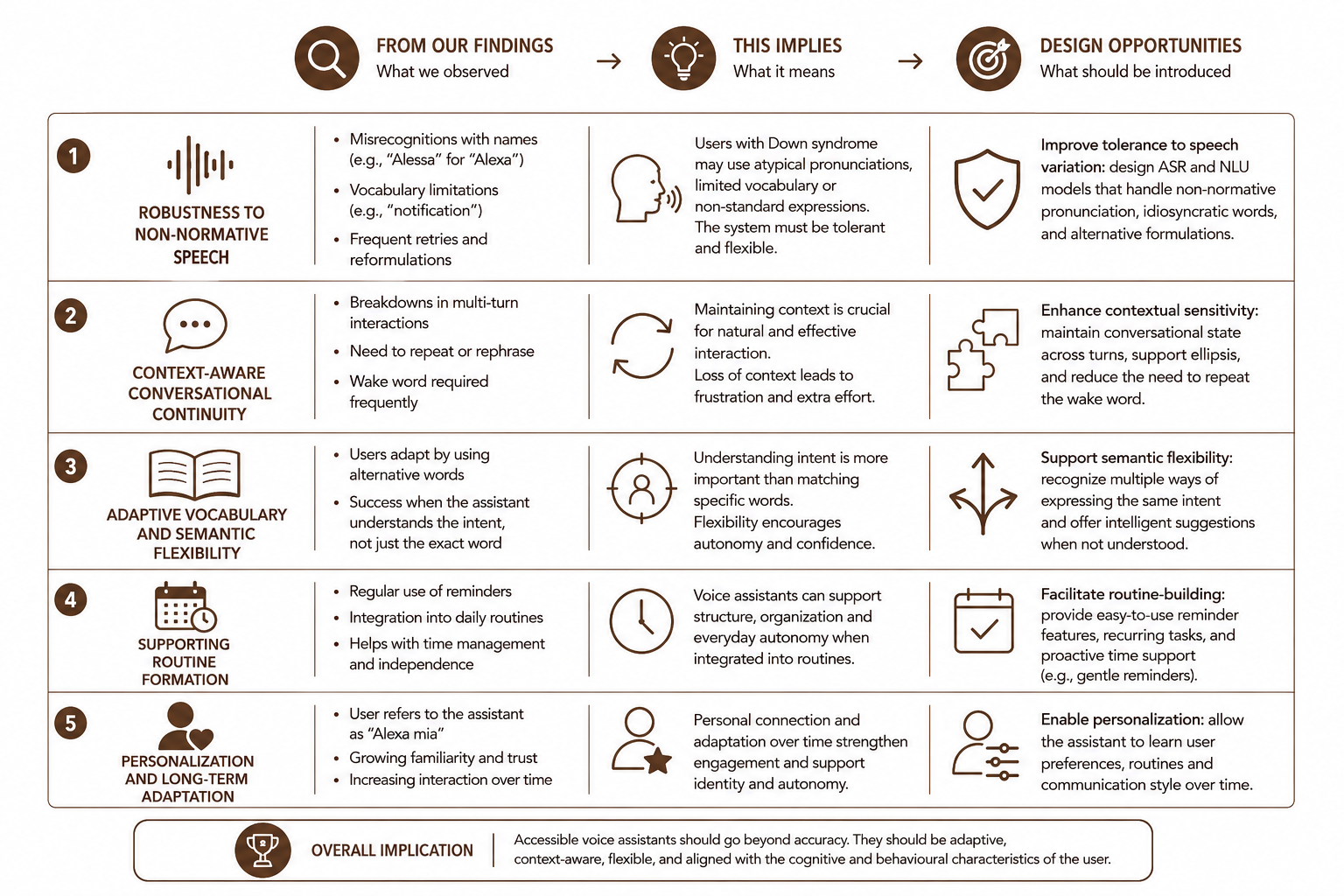}
    \caption{Design implications for accessible voice assistants derived from the longitudinal study. The figure summarizes key findings, the accessibility challenges they reveal, and corresponding opportunities for the design of more adaptive, context-aware, and personalized conversational interfaces.}
    \label{fig:design-implications}
\end{figure*}

Field observations  (resumed in Fig.~\ref{fig:design-implications}) revealed a category of interaction breakdowns that became particularly evident during the later stages of adoption, when the participant began to perceive Alexa not merely as a question-answering system but as a conversational partner. As familiarity with the device increased, interactions progressively shifted from simple commands and requests toward more natural conversational exchanges driven by curiosity and spontaneous follow-up questions.

In these situations, limitations of the conversational interface frequently led to frustration. Unlike human conversations, where pauses and interruptions do not necessarily disrupt the continuity of the interaction, Alexa generally treated each utterance as an independent request. As a consequence, conversational context was often lost between turns. For example, after asking about television programs scheduled for the evening, the participant would sometimes formulate a follow-up question such as `Why is program X not listed?''. Because the follow-up question relied on information established in the previous exchange, Alexa was frequently unable to interpret it correctly and responded with generic messages such as `I don't understand'' or ``I don't know how to answer that''.

A second source of disruption was the need to repeatedly invoke the wake word. While this requirement is functional from a system-design perspective, it introduces a conversational overhead that differs substantially from human dialogue. The repeated need to explicitly re-establish the interaction reduced conversational fluidity and made sustained multi-turn exchanges difficult to maintain.
Interestingly, these difficulties did not emerge during the initial stages of adoption, when interactions primarily consisted of isolated commands and information requests. Instead, they became more apparent as the participant developed greater confidence with the device and attempted to engage in more natural forms of dialogue. This observation suggests that interaction breakdowns may arise not only from accessibility barriers or speech-recognition errors, but also from a mismatch between users' expectations of conversational continuity and the limited contextual capabilities of current voice assistants.

%The observed interaction patterns indicate that accessibility is influenced not only by speech-recognition accuracy but also by conversational continuity, vocabulary flexibility, routine support, and long-term personalization. Based on the quantitative and qualitative findings, we identify a set of design considerations for future voice assistants intended to support users with DS and other cognitive disabilities as illustrated in Fig.~\ref{fig:design-implications}).

Recent advances in Large Language Model (LLM)-based conversational systems suggest a potential solution to many of the interaction breakdowns observed in this study, enabling a transition from intent-based voice assistants toward systems capable of maintaining contextual awareness, supporting multi-turn dialogue, and adapting interactions to users' conversational history \cite{foosherian2023enhancing,acikgoz2025conversational,liu2024raise}. Unlike traditional voice assistants, which typically process each utterance as an independent request, LLM-based conversational systems leverage contextual reasoning and memory to sustain more natural and coherent interactions. These capabilities have recently begun to appear in commercial platforms, including \textit{Amazon Alexa+}~\cite{amazon2025alexaplus},
\textit{Google Gemini}~\cite{google2025gemini},
and \textit{Apple Intelligence}~\cite{apple2025intelligence}, with the aim of improving personalization, conversational continuity, and task completion. 
Such capabilities may be particularly beneficial for users with cognitive disabilities, as they reduce the need to repeatedly invoke the assistant, preserve conversational continuity, and better accommodate spontaneous follow-up questions. 
Although empirical evidence regarding their accessibility for individuals with DS is still largely unavailable, these technological developments suggest promising opportunities to overcome several limitations identified in current voice assistants and establish a baseline for understanding how adults with DS interact with commercial voice assistants before the widespread adoption of LLM-based conversational systems. Future studies should investigate whether the enhanced contextual reasoning, conversational continuity, and personalization offered by next-generation assistants such as Alexa+ translate into measurable improvements in accessibility, communicative engagement, and everyday autonomy

%%=============================================%%
%% For presentation purpose, we have included  %%
%% \bigskip command. Please ignore this.       %%
%%=====================================

%\textbf{Ethical approval declarations} (only required where applicable) Any article reporting experiment/s carried out on (i)~live vertebrate (or higher invertebrates), (ii)~humans or (iii)~human samples must include an unambiguous statement within the methods section that meets the following requirements: 

%\begin{enumerate}[1.]
%\item Approval: a statement which confirms that all experimental protocols were approved by a named institutional and/or licensing committee. Please identify the approving body in the methods section

%\item Accordance: a statement explicitly saying that the methods were carried out in accordance with the relevant guidelines and regulations

%\item Informed consent (for experiments involving humans or human tissue samples): include a statement confirming that informed consent was obtained from all participants and/or their legal guardian/s
%\end{enumerate}

%If your manuscript includes potentially identifying patient/participant information, or if it describes human transplantation research, or if it reports results of a clinical trial then  additional information will be required. Please visit (\url{https://www.nature.com/nature-research/editorial-policies}) for Nature Portfolio journals, (\url{https://www.springer.com/gp/authors-editors/journal-author/journal-author-helpdesk/publishing-ethics/14214}) for Springer Nature journals, or (\url{https://www.biomedcentral.com/getpublished/editorial-policies\#ethics+and+consent}) for BMC.

\section{Conclusion}\label{sec13}

This paper presented a longitudinal case study investigating the everyday use of a commercial voice assistant by an adult with DS over nine months in a natural domestic environment. By combining interaction logs, audio recordings, and qualitative observations, the study provides an ecologically valid account of how conversational technologies are appropriated, integrated into daily routines, and experienced over extended periods of use.

The findings show that voice assistants can support a wide range of everyday activities, including entertainment, information seeking, routine management, and autonomy-supporting practices. Beyond functional use, the results highlight a gradual evolution of the participant's interaction with the system, characterized by communicative adaptation, increasing engagement, and the emergence of socially oriented behaviors such as greetings, anthropomorphic expressions, and curiosity-driven interactions. At the same time, the study identifies important limitations of current commercial voice assistants, including reduced robustness to non-normative speech, dependence on specific lexical formulations, and the inability to maintain conversational context across multiple dialogue turns.

These findings extend current knowledge on accessible conversational interfaces by providing rare longitudinal evidence collected in a real domestic setting. They also suggest a number of design implications for future voice assistants, including greater adaptability to users' linguistic characteristics, more flexible vocabulary handling, improved contextual reasoning, and proactive support for everyday routines. Many of the interaction breakdowns observed in this study appear to stem not from the participant's cognitive abilities but from limitations of current conversational architectures, indicating that recent LLM-based conversational systems may offer promising opportunities to address several of these challenges.

The present work is inherently limited by its single-case design and therefore does not aim to provide statistically generalizable conclusions. Nevertheless, the richness, duration, and ecological validity of the collected data offer analytically transferable insights into long-term human--AI interaction in the context of DS. Future research should extend this approach to larger and more diverse participant groups and investigate how next-generation LLM-based conversational assistants influence accessibility, communicative engagement, and everyday autonomy in real-world settings.

\backmatter

%\bmhead{Supplementary information}

%If your article has accompanying supplementary file/s please state so here. 

%Authors reporting data from electrophoretic gels and blots should supply the full unprocessed scans for key as part of their Supplementary information. This may be requested by the editorial team/s if it is missing.

%Please refer to Journal-level guidance for any specific requirements.

\bmhead{Acknowledgements}
The author gratefully acknowledges the participant and the caregiver, whose sustained engagement made this longitudinal study possible.

\section*{Declarations}
%Some journals require declarations to be submitted in a standardised format. Please check the Instructions for Authors of the journal to which you are submitting to see if you need to complete this section. If yes, your manuscript must contain the following sections under the heading `Declarations':

\begin{itemize}
%\item Funding: Not Applicable
\item \textit{Declaration of generative AI and AI-assisted technologies in the manuscript preparation
process}: During the preparation of this work, the author used AI-assisted coding tools, including
ChatGPT-5.3-codex and ChatGPT-5.5 Edu\footnote{https://openai.com/it-IT/index/introducing-gpt-5-5/}, to support debugging and code development during the data analysis phase. After using these tools, the author reviewed and edited all outputs as needed, and she takes full responsibility for the content of the published article. Generative AI was also used to produce Fig.\ref{fig:usability-challenges} exclusively for graphical rendering. The conceptual design, scientific content, and detailed prompt were entirely developed by the author.
%\item Ethics approval and consent to participate
\item \textit{Author contribution:} C.S. conceived and designed the study, collected, curated, and analyzed the data, interpreted the results, prepared all figures and tables, and wrote, reviewed, and approved the final manuscript.
%\item \textit{Data availability:} Not Applicable
%\item \textit{Materials availability:} Not Applicable
%\item \textit{Code availability:} Not Applicable
\item \textit{Funding:} Not Applicable
\end{itemize}

\noindent
%If any of the sections are not relevant to your manuscript, please include the heading and write `Not applicable' for that section. 

%\bigskip

%\bigskip\noindent
%Springer journals and proceedings: \url{https://www.springer.com/gp/editorial-policies}

%\bigskip\noindent
%Nature Portfolio journals: \url{https://www.nature.com/nature-research/editorial-policies}

%\bigskip\noindent
%\textit{Scientific Reports}: \url{https://www.nature.com/srep/journal-policies/editorial-policies}

%\bigskip\noindent
%BMC journals: %\url{https://www.biomedcentral.com/getpublished/editorial-policies}
%\end{flushleft}

\begin{appendices}




\end{appendices}

\bibliographystyle{sn-basic}
\bibliography{references} 
%\bibliography{references}
%\input sn-article.bbl

\end{document}